\documentclass[mnsc,nonblindrev]{informs4}
\usepackage{siunitx}
\usepackage{mathtools}
\usepackage{amsmath}
\usepackage{setspace}
\usepackage[letterpaper, top=1in, bottom=1in, left=1.00in, right=1.00in]{geometry}
\usepackage{booktabs, array}
\usepackage{multirow}
\usepackage{multicol}
\usepackage{url}
\usepackage{threeparttable}
\OneAndAHalfSpacedXI
\usepackage{natbib}
\usepackage{comment}
\usepackage{verbatim}
\usepackage{graphicx}
\usepackage{eurosym}
\usepackage{amssymb}
\usepackage{amsfonts}
\usepackage{color}
\usepackage{breqn}
\usepackage{hyperref}
\usepackage{lineno}

 \bibpunct[, ]{(}{)}{,}{a}{}{,}%
 \def\bibfont{\small}%
 \def\bibsep{\smallskipamount}%
 \def\bibhang{24pt}%

\definecolor{airforceblue}{rgb}{0.36, 0.54, 0.66}
\definecolor{celestialblue}{rgb}{0.29, 0.59, 0.82}
\definecolor{darkelectricblue}{rgb}{0.33, 0.41, 0.47}
\definecolor{frenchblue}{rgb}{0.0, 0.45, 0.73}

\hypersetup{hidelinks,colorlinks,linkcolor=black,citecolor=black,urlcolor=black}
 
\TheoremsNumberedThrough
\ECRepeatTheorems

\EquationsNumberedThrough    
\MANUSCRIPTNO{}

\makeatletter
\newcommand*\bigcdot{\mathpalette\bigcdot@{.6}}
\newcommand*\bigcdot@[2]{\mathbin{\vcenter{\hbox{\scalebox{#2}{$\m@th#1\bullet$}}}}}
\makeatother

\usepackage[toc,page]{appendix}
\newcolumntype{L}[1]{>{\raggedright\let\newline\\\arraybackslash\hspace{0pt}}m{#1}}
\newcolumntype{C}[1]{>{\centering\let\newline\\\arraybackslash\hspace{0pt}}m{#1}}
\newcolumntype{R}[1]{>{\raggedleft\let\newline\\\arraybackslash\hspace{0pt}}m{#1}}
\newcolumntype{M}{>{\centering\arraybackslash} m{6cm} }

\begin{document}
\RUNAUTHOR{Cui, Ding, Zhu}

\RUNTITLE{Gender Inequality in Research Productivity}

\TITLE{Gender Inequality in Research Productivity during the COVID-19 Pandemic}

\ARTICLEAUTHORS{
\AUTHOR{Ruomeng Cui}\AFF{Goizueta Business School, Emory University, \EMAIL{ruomeng.cui@emory.edu}} 
\AUTHOR{Hao Ding}\AFF{Goizueta Business School, Emory University, \EMAIL{hao.ding@emory.edu}} 
\AUTHOR{Feng Zhu}\AFF{Harvard Business School, Harvard University, \EMAIL{fzhu@hbs.edu}}
}

\ABSTRACT{We study the disproportionate impact of the lockdown as a result of the COVID-19 outbreak on female and male academics' research productivity in social science. The lockdown has caused substantial disruptions to academic activities, requiring people to work from home. How this disruption affects productivity and the related gender equity is an important operations and societal question. We collect data from the largest open-access preprint repository for social science on 41,858 research preprints in 18 disciplines produced by 76,832 authors across 25 countries over a span of two years. We use a difference-in-differences approach leveraging the exogenous pandemic shock. Our results indicate that, in the 10 weeks after the lockdown in the United States, although total research productivity increased by 35 percent, female academics' productivity dropped by 13.2 percent relative to that of male academics. We also show that this intensified productivity gap is more pronounced for assistant professors and for academics in top-ranked universities and is found in six other countries. Our work points out the fairness issue in productivity caused by the lockdown, a finding that universities will find helpful when evaluating faculty productivity. It also helps organizations realize the potential unintended consequences that can arise from telecommuting. 

}

\maketitle

\vspace{-0.1in}

\section{Introduction}

The Coronavirus 2019 (COVID-19) pandemic has significantly changed the way people live and work. Many people have been forced to work from home through telecommuting, potentially affecting their productivity. We study how this pandemic shock affected academics' research productivity using data from the world's largest open-access repository for social science---the Social Science Research Network (SSRN).\footnote{Source:\url{https://en.wikipedia.org/wiki/Social\_Science\_Research\_Network}, accessed June 2020.} We provide evidence that, as a result of the lockdown in the United States, female researchers' productivity dropped significantly relative to that of male researchers.

In response to the pandemic, the US and many other countries required their citizens to stay at home. As a result of the lockdown, many people have had to perform work duties at home along with household duties. Most countries have also closed schools and daycare centers, massively increasing childcare needs. With the childcare provided by grandparents and friends limited due to the social distancing protocol, parents must care for their children themselves. In addition, restaurants have been either closed or do not allow dine-ins, which has increased the need for food preparation at home. Given that women, on average, carry out disproportionately more childcare, domestic labor, and household responsibilities \citep{bianchi2012housework}, they are likely to be more affected than men by the lockdown.

The lockdown has also disrupted how academics carry out their activities. Many countries have closed their universities, so faculty have to teach and conduct research from home. A researcher's productivity is jointly determined by his or her available time for research and research efficiency \citep{kc2020handbook}. First, given the unequal distribution of domestic duties, the pandemic is more likely to burden female researchers with more home-related tasks, leaving them less time to dedicate to research. Second, scientific research generally requires a quiet and interruption-free environment. As a result of the pandemic, female researchers are more likely to multitask between research and home-related tasks and thus to have lower efficiency in conducting research. Together, these factors suggest that female academics' productivity is likely to be disproportionately affected compared with that of male academics.

Anecdotal evidence provides mixed support for this prediction \citep{AJPSSubmission}. A recent survey involving 4,500 principal investigators reported significant and  heterogeneous declines in the time they spend on research \citep{myers2020unequal}. Several journal editors have noticed that, while there was a 20--30-percent increase in submissions after the pandemic started, most was from male academics \citep{reserachlab}. \cite{whoresearch} find a particularly large number of senior male economists, rather than mid-career economists, exploring research questions arising from the COVID-19 shock.  Others have seen no change or have been receiving comparatively more submissions from women since the lockdown \citep{TheLily}. Overall, there is a dearth of systematic evidence on whether and to what extent the shock has affected gender inequality in academia. We provide such systematic evidence, showing an unequal impact on productivity for female and male researchers.

It is an important operations and societal question to understand the change in productivity and the related gender equity caused by the reorganization of work and care at home. In this paper, we use a large dataset on female and male academics' production of new research papers to systematically study whether COVID-19 and the subsequent lockdown have had a disproportionate effect on female academics' productivity. We also identify the academic ranks, universities, and countries in which this inequality is intensified. 

We collect data on all research papers uploaded to SSRN in 18 disciplines from December 2018 to May 2019 and from December 2019 to May 2020. We extract information on paper titles and the authors' names, affiliations, and addresses, which we use to identify the authors' countries and academic ranks, and the ranking of their institutions. We also use their names and faculty webpages to identify their gender. In particular, we use a large database to predict authors' gender. For author names with a prediction confidence level lower than 80 percent, we use Amazon Mechanical Turk to identify gender manually. The final dataset includes 41,858 papers written by 76,832 authors from 25 countries. Our main analysis focuses on US academics; we then perform the same analysis for other countries. 

We use a difference-in-differences (DID) approach to estimate the effect. We compute the number of papers produced by female and male academics in each week, then compare the variations in the productivity gap between genders before and after the start of the lockdown and show that the gap increased after the start of the lockdown. We also validate that female and male authors' preprint volumes followed the same parallel time trend before the lockdown and that there was no significant change in the research productivity gap in 2019 during the same time of year. Taken together, these results suggest that the intensified disparity has primarily been driven by the pandemic shock.

We find that during the 10 weeks since the lockdown began in the US, female academics' research productivity dropped by 13.2 percent compared to that of male academics. The effect persists when we vary the time window since the pandemic outbreak.  Our findings show that when female and male academics face a reorganization of care and work time, women become significantly less productive by producing fewer papers. We also find that the quality of their uploaded papers, measured by the download and view rates, does not change. Finally, we find that the effect is more pronounced among assistant professors and among researchers in top-ranked research universities and that it exists in six other countries. 

While gender inequality has been long documented for academics in terms of tenure evaluation \citep{antecol2018equal}, coauthoring choices \citep{sarsons2017recognition}, and number of citations \citep{ghiasi2015compliance}, the COVID-19 pandemic brings this issue to the forefront. Our study is among the first to rigorously quantify such inequality in research productivity as a result of the pandemic and our results highlight that this disruption has exacerbated gender inequality in the academic world. Because all academics participate together in open competition for promotions and positions, these short-term changes in productivity will affect their long-term career outcomes \citep{pendamicandfemale}. Thus, institutions should take this inequality into consideration when evaluating faculty members.

Our paper contributes to the literature on productivity, a central topic in operations management. Previous studies have examined key determinants of workers' productivity, such as workers' capacity \citep{tan2014does,kc2020handbook}, multitasking \citep{kc2014multi,bray2016multitasking,kc2020heuristic}, peer effects \citep{song2018closing,tan2019you}, and task sequences \citep{ibanez2018discretionary}. We contribute to the literature by showing that the disruption due to the pandemic has significantly enlarged the productivity gap between female and male researchers, highlighting fairness as an important factor in performance evaluation.

Our work also sheds light on the fairness issues that could arise from telecommuting, an operations choice faced by companies. Since working from home can provide a flexible work schedule for employees and reduce office-related costs for companies, an increasing number of companies are choosing this operating model. Between 2005 and 2015, the number of US employees who chose to telecommute increased by 115\% \citep{2005-2015-115}. By 2019, about 16\% of the total workforce in the US was working remotely full time or part time \citep{us-remote16}. During the pandemic, telecommuting is a constraint rather than a choice; many companies were forced to allow telecommuting. But going forward, an increasing number of companies may choose to offer this operating model to provide flexible work schedule to employees and reduce office-related costs. For example, Twitter and Facebook have already announced that their employees could work from home permanently \citep{facebook-twitter} and JP Morgan planned to expand its telecommuting program \citep{jpmorgan}. Despite the growing popularity of telecommuting, scholars and practitioners still lack a comprehensive understanding of the managerial and societal impact of this operational choice  \citep{nicklin2016telecommuting}. We contribute to the literature by pointing out the productivity inequality caused by the lockdown and telecommuting, which might lead to a longer-term unemployment risk for women, an unintended consequence that companies and society should take into account when making their operational choices or designing policies for performance evaluation. Our findings help institutions and firms understand the potential implications in designing and implementing telecommuting.



\section{Literature Review}

Our work is closely related to three streams of literature: productivity, social operations, and telecommuting. 

\subsection{Productivity}
According to the productivity literature \citep{kc2009impact,tan2014does}, working in different environments causes significant changes in operational factors that drive worker's productivity. In our context, due to the pandemic, researchers have to change to working from home, potentially affecting several drivers of productivity identified by research, such as multitasking, workload, task sequence, and peer effects.

Multitasking is particularly relevant to our research context. When working from home during the pandemic, researchers may need to allocate their limited cognitive capacity across home-related and work-related tasks, thus dealing with more distractions arising from multitasking. Prior studies have shown mixed effects of multitasking on workers' productivity, such as an increased service speed with a lower service quality for restaurant waiters \citep{tan2014does} and a slower processing for bank associates \citep{staats2012specialization}. The productivity losses could be greater for jobs requiring greater cognitive capacity. For example, in the judiciary system, reducing multitasking has been shown to help judges focus on the most recent cases, reduce the switching costs between cases, and increase the case completion rate \citep{bray2016multitasking}. In healthcare, excessive multitasking and frequent interruptions in the work flow have been shown to undermine the productivity of discharging \citep{kc2014multi}, processing \citep{berry2017past}, and medication delivery \citep{batt2017early}. 


Workload, task sequences, and peer effects are examples of other operations drivers in researchers' productivity during the pandemic. Workers have been shown to adjust their productivity based on their workload, slowing down when facing more workload and speeding up when facing less because they internalize the congestion cost \citep{kc2009impact,tan2014does}. However, the extra workload could make workers fatigued, which could reduce productivity \citep{salvendy2012handbook, kc2020handbook}. Technology---such as tabletop technology in restaurants---has been shown to improve workers' productivity by reducing their non-value-added tasks, enabling them to focus on more important tasks \citep{tan2020tabletop}. The literature has also identified workers' choice of task sequence as a productivity factor \citep{staats2012specialization,ibanez2018discretionary}. Workers tend to deviate to suboptimal task sequences when facing a higher workload or when fatigued \citep{KC2020task}. Another driver of workers' productivity is peer effects---workers adapt their own productivity to that of their peers' \citep{schultz1998modeling}. For example, having a particularly capable worker on a shift could motivate slower workers to speed up but could also discourage good performers as it becomes more difficult for them to outperform their peers and reach their goals \citep{tan2019you}. Displaying peers' productivity publicly has been shown to improve productivity \citep{song2018closing}. 






\subsection{Social Operations}
This paper sheds light on a key social issue---fairness and equity---in research productivity, adding to the growing literature on the social impact of operational choices. Several recent influential papers by \cite{tang2012research}, \cite{lee2018socially}, and  \cite{dai2020twenty} encourage OM researchers to work on socially responsible topics that are important to corporations and to society at large. Papers have examined the use of review information to reduce racial discrimination arising in the sharing economy \citep{cui2020reducing,mejia2020transparency} and the gender inequality driven by specific compensation schemes \citep{pierce2020peer}. The literature on gender bias has shown evidence that female researchers and students tend to be discredited when they are evaluated alongside equally competent male candidates \citep{moss2012science,sarsons2017recognition}, that women are more likely to be assigned more service-oriented and less desirable tasks \citep{guarino2017faculty} with fewer promotion opportunities \citep{babcock2017gender}, and that women are often responsible for more housework and childcare \citep{schiebinger2010housework,misra2012gender}. 


In our context, when working from home, the unequal distribution of housework means that women are more likely to deal with non-work-related tasks during the lockdown and lose productivity. A recent survey involving 4,500 principal investigators shows that female scientists self-reported a sharper reduction in research time during the lockdown, primarily due to childcare needs \citep{myers2020unequal}. We contribute to the literature by providing evidence that the lockdown affects productivity and exacerbates gender inequity in the workplace, potentially leading to a long-term career risk for women, an unintended consequence that organizations should consider when designing their operations models and performance evaluation policies.

\subsection{Telecommuting}
Our work is also related to the emerging literature on organizations' telecommuting choices. Transitioning from traditional in-office work to telecommuting might affect workers' behavior and productivity through team-work and peer effects. For example,  \cite{dutcher2012does} observe that workers do not indulge in free-riding behavior when a team is made up of in-office workers and telecommuters, and \cite{bloom2015does} demonstrates that telecommuting can improve productivity when it is carried out in a quiet environment. Our work adds to this stream of literature by demonstrating an unexpected social issue of fairness arising from this operating model. 



\section{Theory Development}

A researcher's productivity can be measured as the product of the amount of time he or she can dedicate to research, \textit{Time Available for Research}, and how efficiently he or she conducts research, \textit{Research Efficiency.} This definition is consistent with the key insights from the literature that productivity is determined by two elements: (1) the capacity constraint due to physical or cognitive limitations and (2) efficiency variations with changes in operations factors in the working environment \citep{kc2020handbook}.


For many researchers, the outbreak of COVID-19 has affected their productivity both in terms of time available for research and research efficiency. In response to the pandemic, most countries have closed schools and daycare centers and required that their citizens to be quarantined at home. As a result, researchers from more than 1,100 colleges and universities had to carry out both work and household duties at home \citep{shutdown2020}. We next illustrate how the disruptions change researchers' time available for research and research efficiency separately. We then argue how these changes might be unequal among female and male researchers.

The pandemic changes researchers' working environment, resulting in a need to reallocate their time across research, work-related tasks (such as commuting, social interaction, and service to the academic community), and home-related tasks (such as childcare and house chores). Certain tasks have an in-person nature, such as commuting to the office, serving administrative duties, and interacting socially with colleagues. These tasks are significantly reduced during the lockdown and the time savings could be substantial; for example, workers in the US on average spend an hour commuting each work day \citep{commuting_time}. During the pandemic, researchers could use this additional capacity for research. At the same time, they might have to allocate more time to home-related tasks. Even without the pandemic, working from home constantly exposes researchers to the home environment and home duties. During the lockdown, to make matters worse, many services and amenities such as schools, day-care centers, hospitals, and restaurants have either been closed or operating at a lower capacity. Researchers might need to allocate extra time to domestic duties. For example, childcare could be particularly time-demanding since parents have diminished access to their regular childcare support network, such as professional caregivers, relatives, and friends. Consequently, researchers who are responsible for more house works and childcare are more likely to allocate additional time to domestic duties. In conclusion, a researcher could have either more or less time available due to the pandemic, which depends on the new allocation of home responsibilities. 

Besides the time available for research, the pandemic affects drivers of research efficiency, such as multitasking, task sequence, fatigue, and peer effects. Conducting scientific work often requires hours of interruption-free environment. When working from home, although researchers are not distracted by activities like commuting, administrative duties, and social interactions with colleagues, they are likely to be distracted by childcare and housework, resulting in excessive multitasking.  Multitasking means that workers have to allocate their limited cognitive capacity across multiple tasks. The setup cost associated with switching between tasks and the difficulty of focusing on and organizing relevant information hinder efficiency \citep{kc2014multi}. Multitasking has also been shown to induce stress and frustration \citep{mark2008cost}, make people more easily distracted \citep{levitin2014organized}, and exhaust their cognitive capacity \citep{kc2014multi}. Home duties and urgency often require researchers' immediate attention, forcing them to deviate from their optimal task sequences, which would in turn reduce their research efficiency. Researchers may also encounter a heavier workload from increased housework and thus experience fatigue, which could also reduce efficiency. Last but not least, working from home makes it difficult to observe one's colleagues’ productivity and to discuss research topics with their peers, both of which reduce the positive influence of peers in motivating researchers, inspiring research ideas, and improving  efficiency \citep{song2018closing,tan2019you}. In conclusion, a researcher's research efficiency could go up and down due to the pandemic, depending on the level of impact they get from multitasking, fatigue, altered task sequence, and altered peer effects.




We next demonstrate how the changes in research time and efficiency can be different for women and men. Women are on average, disproportionately burdened with childcare and household responsibilities \citep{bianchi2012housework}. In the US, they are shown to spend almost twice as much time as men on housework and childcare in the US \citep{bianchi2012housework}. Moreover, there are 8.5 million more single mothers than single fathers \citep{alon2020impact}. Even in the gender-egalitarian countries of northern Europe, women are responsible for almost two-thirds of the unpaid work \citep{EUwomenunpaid}. Among heterosexual couples with female breadwinners, women still do most of the care work \citep{chesley2017signs}. The same pattern exists in academia \citep{schiebinger2010housework,andersen2020meta}. Female professors are shown to spend more time doing housework and carework than male professors across various ranks; for example, 34.1 hours per week versus 27.6 hours for lecturers, 29.6 hours per week versus 25.1 hours for assistant professors, and 37.7 hours per week versus 24.5 hours for associate professors \citep{misra2012gender}. 

The lockdown has caused a surge in domestic duties. The unequal distribution of domestic duties means that the pandemic might further enlarge the gap between women and men's domestic workload and thus might affect female and male researchers' productivity unequally. First, in terms of time available for research, female researchers are likely to reallocate more time to domestic duties due to the pandemic than male researchers do, leaving them with less capacity for research. Second, in terms of efficiency, female researchers are more likely to be disrupted by multitasking between research and home-related tasks and consequently deviate from their optimal task sequences.  Taken these factors together, female researchers tend to suffer more from a reduced amount of time available for research and diminished efficiency compared to male researchers, which suggests a disadvantage in women's productivity during the pandemic. We therefore hypothesize that, during the pandemic, female researchers are more likely to be disproportionately affected in their productivity compared with male researchers.

\section{Data and Summary Statistics}
We collect data from the Social Science Research Network (SSRN), a repository of preprints with the objective of rapidly disseminating scholarly research in social science. We gather data on \emph{all} social science preprints submitted from December 2018 to May 2019 and from December 2019 to May 2020. We extract information on paper titles and the authors' names, affiliations, and addresses. We use the authors' addresses to identify their countries. The COVID-19 outbreak began at different times in different countries, so we collect each country's lockdown start date from news sources and a United Nations report.\footnote{Source:\url{https://en.unesco.org/covid19/educationresponse}, accessed June 2020.} We drop authors without addresses or with addresses in more than one country because we cannot determine when these authors were affected by the lockdown. We also drop countries with fewer than 800 submissions during our study period. The final data set consists of a total of 41,858 papers in 18 disciplines produced by 76,832 authors from 25 countries.

To identify the authors' genders, we first use a database called \textit{Genderize},\footnote{Source:\url{https://genderize.io/}, accessed June 2020.} which predicts gender based on first name and provides a confidence level. About 78 percent of the authors' genders were identified with confidence levels over 80 percent. For the remaining authors, we use Amazon Mechanical Turk to manually search for their professional webpages based on names and affiliations and 
then infer their genders from their profile photos. Our dataset contains 21,733 female academics and 55,099 male academics.

We aggregate the number of new preprints at the weekly level. We then count the number of papers uploaded by each author in each week. To measure the \emph{effective} productivity for preprints with multiple authors, when a preprint has $n$ authors,\footnote{Note that in many social science disciplines, author names are listed in alphabetical order.} each author gets a publication count of $1/n$.\footnote{Note that the validity of this measure relies on the assumption that female and male researchers' relative contributions to the paper do not change significantly after the lockdown. If female researchers have decreased their contributions to the teamwork since the lockdown, this measure would underestimate gender inequality during the pandemic. In Section~7.4, we study \textit{all-male} and \textit{all-female} preprints as an alternative measure, which minimizes potential work shifting across genders.} Finally, we aggregate the effective number of papers to the gender level: in each week, we count the total number of papers produced by male and female authors separately in each social science discipline.

\begin{figure}[h]
\caption{Time Trends of US Preprints from December 2019 to May 2020}
\label{fig:trend2020}
\vspace{0.05in}
\centering
\includegraphics[scale=0.38]{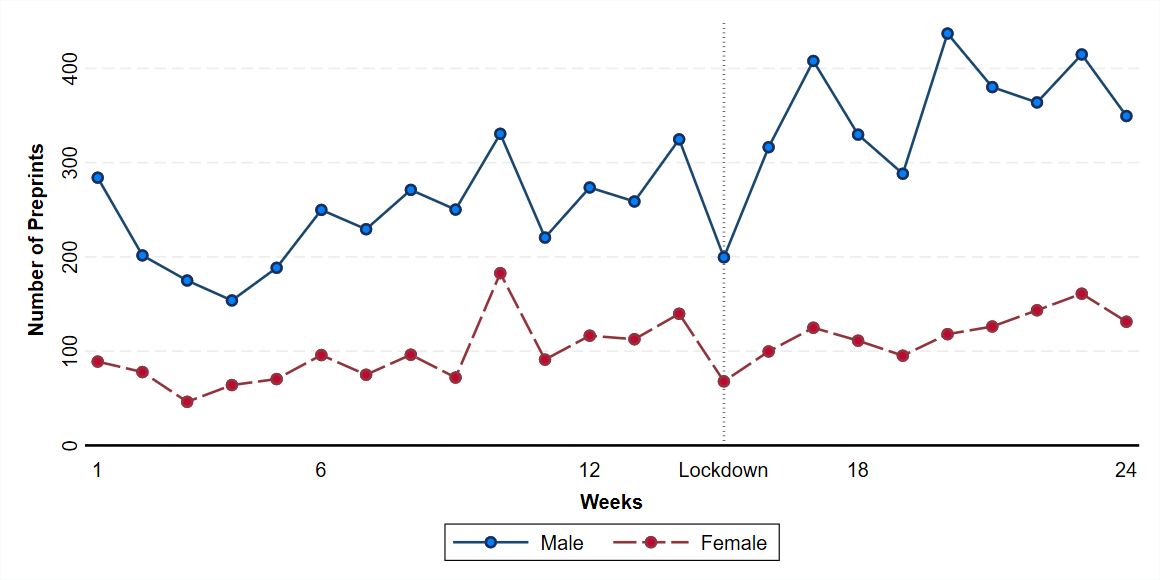}
\vspace{0.05in}

\noindent \scriptsize This graph plots the time trend of the number of preprints for female and male academics. The vertical line represents the start of the lockdown due to COVID-19 in the US.
\end{figure}

Figure \ref{fig:trend2020} plots the time trend of preprints in aggregation from December 3, 2019 to May 19, 2020 in the US. The vertical line represents the week of March 11, 2020, on which nationwide lockdown measures began in the US.\footnote{Most universities were closed in the week of March 11, 2020. Source:  \url{https://gist.github.com/jessejanderson/09155afe313914498a32baa477584fae?from=singlemessage&isappinstalled=0}, accessed June 2020.} We can observe that male academics, on average, have submitted more preprints than female academics, and that female and male academics' research productivity evolved in parallel before the lockdown. After the lockdown started, however, male academics significantly boosted their productivity, whereas female academics' productivity did not change much, indicating an increased productivity gap. 

To ensure that our results are not driven by seasonality, we plot the time trend of preprints during the same time window in 2019 in Appendix Figure \ref{fig:trend2019}. We observe a similar pattern before the week of March 11, 2019, but there is no significant change in the productivity gap after that week. 

We use authors' professional information to identify academic ranks (e.g., PhD student or assistant, associate or full professor). Specifically, we ask workers on Amazon Mechanical Turk to find each researcher's curriculum vitae or professional webpage. To study how the pandemic affects researchers with different academic ranks, we categorize researchers into \textit{students} (which includes PhD and postdoctoral students), \textit{assistant professors}, \textit{associate professors}, and \textit{full professors}, each accounting for 20.3 percent, 16.1 percent, 19.7 percent, and 44.0 percent of the population, respectively. Next, we use authors' affiliations to classify the ranking of their universities. To ascertain whether the productivity gap is intensified or weakened across top-ranked and lower-ranked research universities, we collect social science research rankings from three sources: QS University Ranking,\footnote{Source:\url{https://www.topuniversities.com/university-rankings/university-subject-rankings/2020/social-sciences-management}, accessed June 2020.} Times Higher Education,\footnote{Source:\url{https://www.timeshighereducation.com/world-university-rankings/2020/subject-ranking/social-sciences\#!/page/0/length/25/sort_by/rank/sort_order/asc/cols/stats}, 
accessed June 2020.} and Academic Ranking of World University.\footnote{Source:\url{http://www.shanghairanking.com/FieldSOC2016.html}, accessed June 2020.} We then use these data to rank US universities in order to study the heterogeneous effects across university rankings.

Table~\ref{table:statistics} reports the summary statistics for the weekly number of preprints by gender and discipline from December 3, 2019 to May 19, 2020, spanning 24 weeks, as well as split-sample statistics prior to or after the lockdown. This sample includes 9,943 preprints produced by 15,494 authors in the US and 21,065 preprints produced by 37,997 authors across all countries. The average number of submissions per week is 444.6 in the US and 877.7 across all 25 countries. Notably, while research productivity in the US increased by 35 percent after the lockdown, male authors seem to be the main contributors to this increase. 

About 78 percent of the preprints fall under multiple disciplines.\footnote{Authors self-classify their own preprints into disciplines when they upload their papers. SSRN reviews and approves these classifications.} Note that when computing the total preprints, we count the paper only once when aggregating across disciplines to avoid multiple counting. When computing the number of preprints in each discipline, we separately count all of the papers classified under each one. We observe substantial variations across disciplines. Among the 18 disciplines, Political Science, Economics, and Law received the most submissions, whereas Geography, Criminal Justice and Education the fewest. While there was a large increase in productivity in several disciplines, such as Economics, Political Science, Finance, Health Economics, and  Sustainability, after the COVID-19 outbreak, other disciplines showed no obvious increase. A few disciplines, such as Anthropology, Cognitive, and Information Systems, even experienced a decline.

\begin{table}[h]\centering \scriptsize
    \caption{Summary Statistics} 
    \label{table:statistics}
    \begin{threeparttable}
    \begin{tabular}{C{1.4cm} l*{9}{S[table-format=4.2, table-number-alignment=center]}}

\toprule
 &  &  \multicolumn{5}{c}{All observations} & \multicolumn{2}{c}{Before lockdown} & \multicolumn{2}{c}{After lockdown} \\
 \cmidrule(lr){3-7} \cmidrule(lr){8-9} \cmidrule(lr){10-11} 

Level  &  \centering{Weekly No. of Preprints} &  {Mean}  & {Std. dev}  & {Max} & {Min} & {Total} &  {Mean}  & {Std. dev} &  {Mean}  & {Std. dev} \\

\midrule

\multirow{3}{1.4cm}{\centering All Disciplines (US only)} 
& All  & 444.6 & 109.4 & 617 & 224 & {9,934} & 378.8 & 88.0 & 511.4 & 86.0 \\
& Female authors & 111.3 & 30.8 & 186 & 47 &  {2,493}  & 103.4 & 36.2 & 119.3 & 21.4 \\
& Male authors  & 333.3 & 85.3 & 480 & 161 & {7,441} & 275.4 & 55.4 & 392.1 & 68.6\\
\hline \\[-1.8ex]

\multirow{18}{1.4cm}{\centering By Discipline (US only)} 
& Accounting    & 19.5 & 7.2 & 40 & 9 & {468} & 17.9 & 6.3 & 21.8 & 8.2 \\
& Anthropology  & 85.0 & 21.5 & 141 & 63 & {2,040} & 93.9 & 24.0 & 72.5 & 6.9\\
& Cognitive    & 11.3 & 9.2 & 31 & 1 & {271} & 14.1 & 11.1 & 7.4 & 3.2 \\
& Corporate & 14.1 & 6.5 & 27 & 3 & {339} & 12.2 & 6.5 & 16.8 & 5.8 \\
& Criminal & 15.4 & 6.7 & 27 & 4 & {370} & 12.8 & 6.7 & 19.1 & 4.9 \\
& Economics & 133.2 & 54.2 & 237 & 37 & {3,197} & 106.6 & 39.1 & 170.5 & 51.6 \\
& Education & 17.9 & 7.0 & 36 & 7 & {429} & 16.9 & 7.4 & 19.2 & 6.7 \\
& Entrepreneurship & 9.9 & 5.3 & 22 & 2 & {238} & 10.2 & 4.9 & 9.5 & 5.9 \\
& Finance & 91.7 & 34.5 & 139 & 25 & {2,201} & 78.5 & 35.5 & 110.2 & 24.0 \\
& Geography & 8.2 & 3.3 & 17 & 3 & {196} & 7.5 & 2.7 & 9.1 & 4.0 \\
& Health Economics & 8.4 & 10.1 & 47 & 0 & {202} & 3.0 & 2.1 & 16.0 & 12.1 \\
& Information Systems & 15.6 & 7.3 & 39 & 7 & {374} & 17.4 & 8.6 & 13.1 & 4.2 \\
& Law & 98.5 & 24.3 & 142 & 44 & {2,365} & 94.1 & 26.7 & 104.7 & 20.1 \\
& Management & 33.4 & 11.4 & 56 & 12 & {802} & 33.4 & 13.3 & 33.4 & 8.6 \\
& Organization & 20.5 & 11.5 & 44 & 3 & {491} & 16.9 & 10.2 & 25.5 & 11.7 \\
& Political Science & 167.9 & 50.5 & 255 & 85 & {4,030} & 142.1 & 39.0 & 204.1 & 42.8 \\
& Sustainability & 22.8 & 11.9 & 66 & 8 & {546} & 18.1 & 5.9 & 29.3 & 15.1 \\
& Women/Gender & 18.0 & 4.7 & 28 & 10 & {431} & 17.2 & 4.4 & 19.0 & 5.2 \\

\hline \\[-1.8ex]

\multirow{3}{1.4cm}{\centering All countries}
& All & 877.7 & 199.3 & {1,175} & 487 & {21,065} & 779.1 & 177.5 & 1015.8 & 140.4 \\
& Female authors & 246.5 & 53.9 & {347} & 165 & {5,916} & 231.0 & 57.0 & 268.2 & 42.9 \\
& Male authors  & 631.2 & 152.0 & {866} & 322 & {15,149}  & 548.1 & 124.4 & 747.6 & 104.3 \\

\bottomrule
\end{tabular}

\noindent The table summarizes the weekly number of papers from December 2019 to May 2020. The sample includes 15,494 authors from the United States and 37,997 authors across all countries. There are 9,934  preprints produced by US authors, 2,493 of which are produced by 3,877 female researchers and 7,441 by 11,617 male researchers. We gather the country-specific lockdown time to split our sample to before and after the lockdown for each country.

\end{threeparttable}
\end{table}

\section{Identification}

Our identification exploits the lockdown in response to the COVID-19 outbreak as an exogenous shock that has caused substantial disruption to academic activities, requiring academics to teach, conduct research, and carry out household duties at home. The validity of our approach depends on the assumption that the shock is exogenous with respect to the researchers' anticipated responses. If a particular gender group of researchers anticipated and strategically prepared for the shock by accelerating the completion of their research papers, this could confound the treatment effect. In reality, this possibility is unlikely because of the rapid development of the situation. COVID-19 was regarded as a low risk and not a threat to the US in late January \citep{fauci}, and no significant actions had been taken other than travel warnings issued until late February \citep{travelban}. It quickly turned into a global pandemic after the declaration of the World Health Organization on March 11, 2020, followed by the nationwide shelter-in-place orders within a week.\footnote{Source:\url{https://www.cdc.gov/nchs/data/icd/Announcement-New-ICD-code-for-coronavirus-3-18-2020.pdf}, accessed June 2020.} 

We adopt the difference-in-differences (DID) methodology, a common approach used to evaluate people's or organizations' responses to natural shocks \citep[e.g.,][]{seamans2013responses,wagner2018disclosing}. We perform the DID analysis using outcome variables on two levels: the number of preprints in each discipline and the number of preprints aggregated across all disciplines.\footnote{We also perform a DID analysis with the country-level panel data. For this, we assign a lockdown dummy to each country and combine data across countries to form a country-level panel. This analysis yields consistent results.}

We compare the productivity gap between female and male researchers before and after the pandemic outbreak using the following model specification with discipline-level panel data:

\begin{equation}\label{eq:did_discipline}
log(Preprints_{igt}) = c + Female_{g}  + \beta Female_{g}\times Lockdown_{t} + \gamma_t + \delta_i + \epsilon_{igt},
\end{equation}
where $i$ denotes discipline, $g$ denotes the gender group, $t$ denotes the week, $log(Preprints_{igt})$ represents the logged number of preprints uploaded for discipline $i$ for gender $g$ during week $t$, $\gamma_t$ is the time fixed effect, $\delta_i$ is the discipline fixed effect that captures the time-invariant characteristics of discipline $i$, and $\epsilon_{igt}$ is the error term. The time fixed effect $\gamma_t$ includes weekly time dummies that control for time trends. The dummy variable $Female_g$ equals 1 if gender $g$ is female, and 0 otherwise. The dummy variable $Lockdown_{t}$ equals 1 if week $t$ occurs 
after the lockdown measure was adopted (that is, the week of March 11, 2020), and 0 otherwise. Its main effect is absorbed by the time fixed effects. The coefficient $\beta$ estimates the effect of the lockdown on female academics' research productivity relative to male academics’ productivity. 

We also use aggregate-level data to estimate the effect with the following DID specification:
\begin{equation}\label{eq:did}
log(Preprints_{gt}) = c + Female_{g}  + \beta Female_{g} \times Lockdown_{t} + \gamma_t + \epsilon_{gt},
\end{equation}
where $g$ denotes the gender group, $t$ denotes the week, $log(Preprints_{gt})$ represents the logged number of preprints uploaded for gender $g$ during week $t$, and $\epsilon_{t}$ is the error term. As before, we include the time fixed effect $\gamma_t$.

\section{Empirical Results}

In this section, we report the effect of the COVID-19 outbreak on research productivity. We first show the average effect of the pandemic on gender inequality. We then show the heterogeneous effects across academic disciplines, faculty ranks, university rankings, and countries.

\subsection{Main Results}

Table~\ref{table:did_discipline} reports the estimates with the discipline fixed effect using Equation~(\ref{eq:did_discipline}). Table~\ref{table:did_aggregated} reports the estimated effect of the pandemic shock on research productivity at the aggregate level using Equation~(\ref{eq:did}). In each analysis, we use 14 weeks before the lockdown as the pre-treatment period and 6 to 10 weeks after the lockdown as the post-treatment periods. The analyses yield consistent results. First, in line with our summary statistics, the results show that fewer preprints are produced by female academics than by male academics in general. Second, since the lockdown began, there has been a significant reduction in female academics' productivity relative to that of their male colleagues, indicating an exacerbated productivity gap in gender. 
The coefficient of the interaction term in Column (5) suggests a 13.2-percent reduction in females' productivity over the 10-week period after the lockdown relative to the males'.\footnote{Because the outcome variable is logged, the percentage change in the outcome variable is computed as $e^{\text{coefficient}}-1$.}

\renewcommand{\arraystretch}{1.1}
\begin{table}[h] \centering \scriptsize \vspace{-0.1in}
   \caption{Impact of Lockdown on Gender Inequality with the Discipline Fixed Effect}
 \label{table:did_discipline}
 \begin{threeparttable}
 \begin{tabular}{ L{4cm} @{}c*{5}{S[table-format=-1.3,table-space-text-post=***,table-column-width=1.8cm]}@{}}
\toprule
 & \multicolumn{5}{c}{Dependent variable: No. of Preprints (in logarithm) by discipline} \\
 
 & {6 weeks} & {7 weeks} & {8 weeks} & {9 weeks} & {10 weeks} \\
 
{Variables} & {(1)} & {(2)} & {(3)} & {(4)}  & {(5)}  \\

\hline \\[-1.8ex]

{$Female$}  & $-$0.791*** & -0.791***  & -0.791*** & -0.791*** & -0.791*** \\
  & {(0.042)} & {(0.042)} & {(0.042)} & {(0.042)} & {(0.042)} \\ 
  
{$Female \times Lockdown$} &$-$0.140* &  -0.148**  & -0.162**  & -0.157**  & -0.142** \\
  & {(0.076)} & {(0.072)} & {(0.068)} & {(0.065)} & {(0.063)} \\
 
Discipline fixed effects  & {Yes} & {Yes} & {Yes} & {Yes} & {Yes} \\
Time fixed effects & {Yes} & {Yes} & {Yes} & {Yes} & {Yes} \\
Observations & {720} & {756} & {792} & {828} & {864} \\
$R^{2}$ & 0.837 & 0.836 & 0.839 & 0.841 & 0.841 \\
\bottomrule 
\end{tabular}
\noindent This table reports the estimated coefficients and robust standard errors (in parentheses) in Equation~\eqref{eq:did_discipline} with the discipline fixed effect. The coefficients for 6, 7, 8, 9, and 10 weeks since the lockdown are presented in Columns (1)--(5), respectively.  Significance at $^{*}p<0.1$; $^{**}p<0.05$; $^{***}p < 0.01$.
\end{threeparttable}
\end{table}

\renewcommand{\arraystretch}{1.1}
\begin{table}[h] \centering \scriptsize \vspace{-0.1in}
 \caption{Impact of Lockdown on Gender Inequality in Aggregation}
 \label{table:did_aggregated}
 \begin{threeparttable}
 \begin{tabular}{L{4cm} @{}c*{5}{S[table-format=-1.3,table-space-text-post=***,table-column-width=1.8cm]}@{}}
\toprule
 & \multicolumn{5}{c}{Dependent variable: No. of Preprints (in logarithm) in aggregation} \\
 
 & {6 weeks} & {7 weeks} & {8 weeks} & {9 weeks} & {10 weeks} \\

{Variables} & {(1)} & {(2)} & {(3)} & {(4)}  & {(5)}  \\

\hline \\[-1.8ex]

{$Female$}  & $-$1.013*** & -1.013***  & -1.013*** & -1.013*** & -1.013*** \\
  & {(0.054)} & {(0.054)} & {(0.053)} & {(0.053)} & {(0.053)} \\ 
  
{$Female \times Lockdown$} & $-$0.197** & -0.199***  & -0.173** & -0.159** & -0.150** \\
  & {(0.068)} & {(0.064)} & {(0.067)} & {(0.066)} & {(0.064)} \\
  
{Time fixed effects} & {Yes} & {Yes} & {Yes} & {Yes} & {Yes} \\
Observations & {40} & {42} & {44} & {46} & {48} \\
$R^{2}$ & 0.981 & 0.982 & 0.982 & 0.982 & 0.983 \\
\bottomrule 
\end{tabular}
\noindent This table reports the estimated coefficients and robust standard errors (in parentheses) in Equation~\eqref{eq:did} at the aggregation level. The coefficients for 6, 7, 8, 9, and 10 weeks since the lockdown are presented in Columns (1)--(5), respectively. Significance at $^{*}p<0.1$; $^{**}p<0.05$; $^{***}p < 0.01$.
\end{threeparttable}
\end{table} 

\subsection{Heterogeneous Effects}


In this section, we study the heterogeneous effects across faculty ranks, university rankings, disciplines, and countries.

\subsubsection{Effects across academic rank.}

We study how the pandemic affects researchers with different academic ranks (such as PhD student, or assistant, associate or full professor).  Because assistant professors often face more pressure than senior professors to publish papers in order to get tenure, they are more motivated to devote as much time as possible to research. They are also at a stage at which many have young children. As a result, the pandemic's effect on the productivity gap is likely to be more pronounced for this group.  We repeat the DID analysis for each academic rank and report results based on Equations~\eqref{eq:did_discipline} and~\eqref{eq:did} in Table~\ref{table:rank_discipline} and Table~\ref{table:rank_aggregated}, respectively. The two tables show consistent results. Table~\ref{table:rank_discipline} shows that female assistant professors experienced the most significant drop in research productivity (compared to male junior faculty) since the lockdown. 

\renewcommand{\arraystretch}{1.1}
\begin{table}[h] \centering \scriptsize \vspace{-0.1in}
   \caption{Impact of Lockdown on Gender Inequality by Academic Ranks}
 \label{table:rank_discipline}
 \begin{threeparttable}
\begin{tabular}{  @{}l*{5}{S[table-format=-1.3,table-space-text-post=***,table-column-width=1.9cm]}@{}}
\toprule

 & \multicolumn{5}{c}{Dependent variable: No. of Preprints (in logarithm) by discipline} \\

 & {6 weeks} & {7 weeks} & {8 weeks} & {9 weeks} & {10 weeks} \\
 
{Academic rank} & {(1)} & {(2)} & {(3)} & {(4)}  & {(5)}  \\

\hline \\[-1.8ex]

Student & -0.002 &  -0.016  & -0.014  & -0.038  & -0.071 \\
Assistant professor & -0.529*** & -0.419*** & -0.448*** & -0.419*** & -0.441*** \\
Associate professor & 0.038 & -0.026 & 0.004 & -0.044 & -0.034 \\
Full professor & -0.181 & -0.163 & -0.063 & -0.044 & -0.064 \\

Observations & {720} & {756} & {792} & {828} & {864} \\
\bottomrule 
\end{tabular}
\noindent  This table reports the estimated coefficients based on Equation~\eqref{eq:did_discipline} with the discipline fixed effect for academics within each rank. The coefficients for 6, 7, 8, 9, and 10 weeks since the lockdown are presented in Columns (1)--(5), respectively. Time and discipline fixed effects at the weekly level are included in each regression. Standard errors and estimates of other variables are omitted for brevity. Significance at $^{*}p<0.1$; $^{**}p<0.05$; $^{***}p < 0.01$.
\end{threeparttable}
\end{table}

\subsubsection{Effects across university ranking.} Table~\ref{table:ranking} replicates the DID analysis using Equation~(\ref{eq:did_discipline}) for a subset of academics based on the rankings of their universities.\footnote{For authors affiliated with more than one academic institutions, we use the highest-ranked institution.} Due to our focus on social science, we use the 2020 QS World University Ranking for social sciences and management for the main analysis. We separately analyze academics in universities ranked in the top 10, 20,..., and 100. The results show that the COVID-19 effect is more pronounced in top-tier universities and that this effect in general decreases and becomes less significant as we include more lower-ranked universities. We find similar results when using the two other rankings, as shown in Appendix Table~\ref{table:rankingrobust}.

\renewcommand{\arraystretch}{1.1}
\begin{table}[h] \centering \scriptsize \vspace{-0.1in}
   \caption{Impact of Lockdown on Gender Inequality by University Ranking}
 \label{table:ranking}
 \begin{threeparttable}
\begin{tabular}{  @{}l*{5}{S[table-format=-1.3,table-space-text-post=***,table-column-width=1.8cm]}@{}}
\toprule

 & \multicolumn{5}{c}{Dependent variable: No. of Preprints (in logarithm) by discipline} \\

{Universities} & {6 weeks} & {7 weeks} & {8 weeks} & {9 weeks} & {10 weeks} \\
 
{by QS Ranking} & {(1)} & {(2)} & {(3)} & {(4)}  & {(5)}  \\

\hline \\[-1.8ex]

Top 10 & -0.169** & -0.199*** & -0.158** & -0.153** & -0.165** \\
Top 20 & -0.181** & -0.215*** & -0.183** & -0.179*** & -0.183*** \\
Top 30 & -0.189** & -0.210** & -0.167** & -0.168** & -0.170** \\
Top 40 & -0.218*** & -0.238*** & -0.200*** & -0.191*** & -0.194*** \\
Top 50 & -0.197** & -0.214*** & -0.180*** & -0.179*** & -0.182*** \\
Top 60 & -0.138* & -0.163* & -0.145* & -0.143** & -0.155** \\
Top 70 & -0.142* & -0.155* & -0.132* & -0.122* & -0.127* \\
Top 80 & -0.139* & -0.149** & -0.130* & -0.123* & -0.126* \\
Top 90 & -0.118 & -0.124* & -0.101 & -0.097 & -0.097 \\
Top 100 & -0.100 & -0.102 & -0.083 & -0.082 & -0.090 \\
 
Observations & {720} & {756} & {792} & {828} & {864} \\
\bottomrule 
\end{tabular}
\noindent  This table reports the estimated coefficients based on Equation~\eqref{eq:did} at the aggregate level for universities within each rank group. The coefficients for 6, 7, 8, 9, and 10 weeks since the lockdown are presented in Columns (1)--(5), respectively. Time fixed effects at the weekly level are included in each regression. Standard errors and estimates of other variables are omitted for brevity. Significance at $^{*}p<0.1$; $^{**}p<0.05$; $^{***}p < 0.01$.
\end{threeparttable}
\end{table} 

As junior faculty, researchers in higher-ranked universities often face more pressure to publish papers in order to get tenure and therefore devote more time to research. The constraint caused by the pandemic therefore has a bigger impact on researchers from top-tier universities, resulting in a greater gender inequality among them.

\subsubsection{Effects across countries.} 
\begin{figure}[h]
\caption{Impact of Lockdown on Gender Inequality across Countries}
\label{fig:countries}
\vspace{0.05in}
\centering
\includegraphics[scale=0.36]{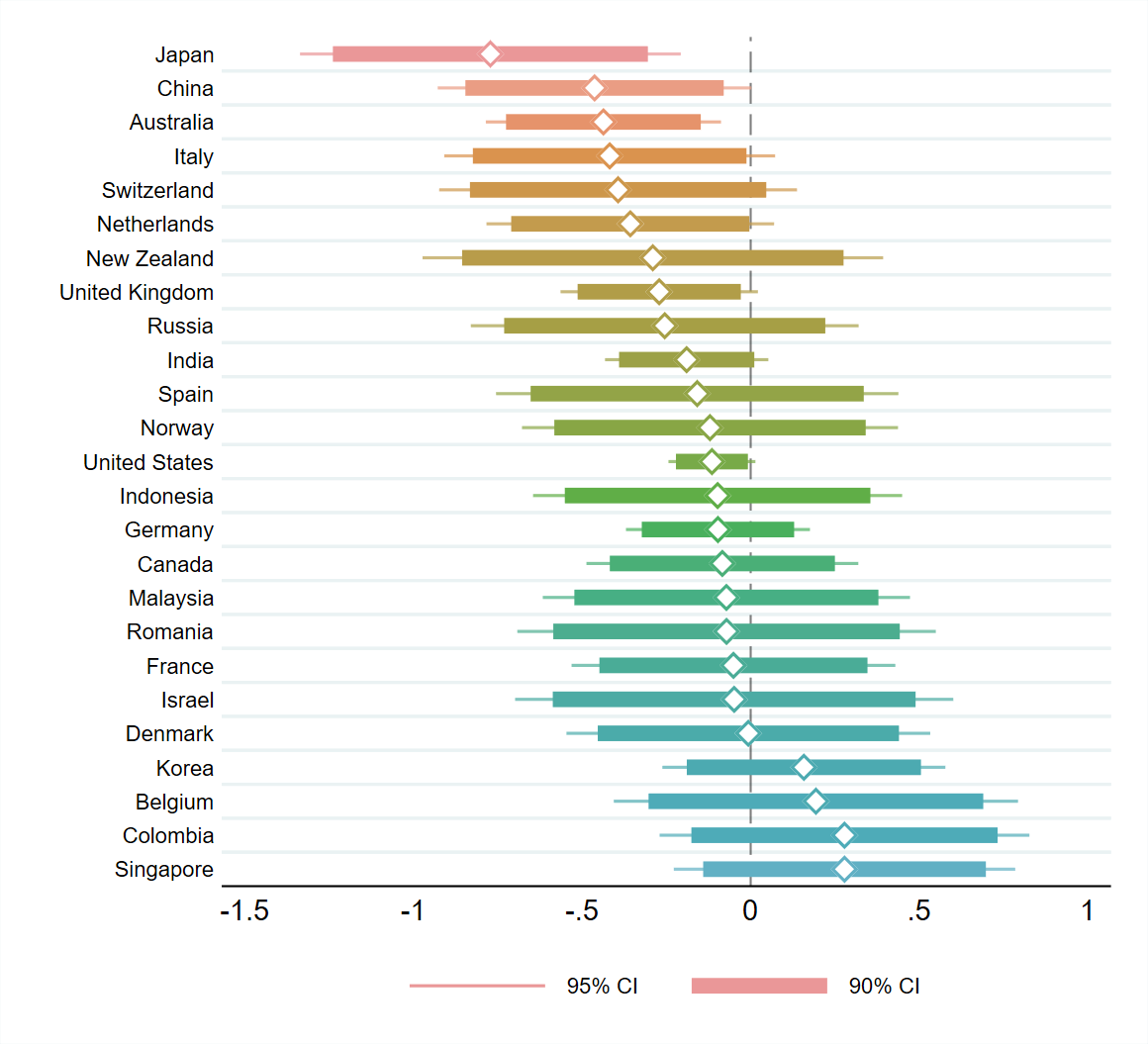}
\vspace{0.05in}

\noindent \scriptsize This graph plots the estimates of the interacted term with 90-percent and 95-percent confidence intervals in each country. The negative values represent female academics' research productivity drop relative to that of male academics. 
\end{figure}

Finally, we examine how the estimated gender inequality varies across countries by replicating the analysis for academics in each country. Figure~\ref{fig:countries} illustrates the impact on the productivity gap graphically by plotting the estimates of the interacted term with 90-percent and 95-percent confidence intervals; a negative value represents a drop in female academics' research productivity relative to that of male academics. We can observe that most countries---21 out of 25 countries---have experienced a decline in female researchers' productivity. In addition to the US, six countries have shown statistically significant declines: Japan, China, Australia, Italy, the Netherlands, and the United Kingdom. Note that because SSRN is primarily used by US researchers, its preprints for other countries might be limited, which might weaken our ability to detect changes.

\section{Robustness Checks}

In this section, we report the results of several robustness tests. Specifically, we check the parallel trends assumption and conduct falsification tests to ensure that our estimated effects are not idiosyncratic. We also test gender inequality in research quality, examine the co-authorship issue, and rule out the data censoring concern.

\subsection{Parallel trends}

The key identification assumption for the DID estimation is the parallel trends assumption; namely, that before the COVID-19 shock, female and male researchers' productivity would follow the same time trend. In Appendix Figure~\ref{fig:trend2019}, which presents the time trends of preprints in 2019, visual inspection shows the two genders' parallel evolution before the shock. We then test this assumption by performing an analysis similar to that of \cite{seamans2013responses} and \cite{wagner2018disclosing}, in which we expand Equation~\eqref{eq:did_discipline} to estimate the treatment effect week by week before the shock. Specifically, we replace $Lockdown_{t}$ in Equation~\eqref{eq:did}  with the dummy variable $Time^{t}_{\tau}$,  where $\tau\in\{-14,-13,...,-2,-1,0\}$ and $Time^{t}_{\tau}=1$ if $\tau=t$ and 0 otherwise, indicating the relative $\tau$th week of the outbreak,
\begin{equation}\label{eq:did_parallel}
log(Preprint_{igt}) = c + Female_{g} + \sum^{-1}_{\tau=-14} Time^t_{\tau} + \sum^{-1}_{\tau=-14} \beta_{\tau} Female_{g}\times Time^t_{\tau} + \delta_i + \epsilon_{igt}.
\end{equation}
The benchmark group is the week of the pandemic outbreak. The coefficients $\beta_{-14}$ to $\beta_{-1}$ identify any week-by-week pre-treatment difference between the female and male researchers, which we expect to be insignificant. We then repeat the same analysis with our aggregate-level data. 

Appendix Table \ref{table:trend} presents the estimation results. They show no pre-treatment differences in the research productivity trends of female and male academics, which supports the parallel trends assumption.

\subsection{Falsification test}

To show that our estimate effects are not an artifact of seasonality, we test whether such a decline in female productivity also occurred in 2019. Appendix Table~\ref{table:statistics2} reports the summary statistics in 2019. We repeat the same analysis specified in Equation~(\ref{eq:did}) using data in 2019 for the same time window used in 2020. If our results simply capture seasonality,  we would be able to find significant effects in 2019. Appendix Table \ref{table:falsification} reports the falsification test results. The placebo-treated average treatment effects are insignificant, implying that women's productivity did not decline significantly in the previous year.

\subsection{Research Quality} So far, we study the research quantity--- using \textit{Number of Preprints} as the dependent variable. One might question whether
the difference in productivity is because, since the lockdown began, male researchers increased the volume of their production at the expense of quality. If this is true, the relative quality of female researchers' preprints should have increased since the lockdown. We test this possibility using data on how many times the abstract has been viewed and the preprint has been downloaded for each preprint, the two primary quality indicators used by SSRN to rank preprints. Appendix Table~\ref{table:statistics_downloads_views} reports the summary statistics of these two variables. We compare the average number of abstract views per preprint and the average number of downloads per preprint for preprints from male and female researchers prior to and after the pandemic outbreak using the same specification as in Equation~(\ref{eq:did_discipline}) with the discipline fixed effect:
\begin{equation}\label{eq:did_view_download}
log(Abstract\ Views_{gt}~or~Downloads_{gt}) = c + Female_{g}  + \beta Female_{g}\times Lockdown_{t} + \gamma_t + \delta_i + \epsilon_{igt}.
\end{equation}

\renewcommand{\arraystretch}{1.1}
\begin{table}[h] \centering \scriptsize \vspace{-0.1in}
 \caption{Impact of Lockdown on Abstract Views}
 \label{table:views_displine}
 \begin{threeparttable}
 \begin{tabular}{L{4cm} @{}l*{5}{S[table-format=-1.3,table-space-text-post=***,table-column-width=2cm]}@{}}
\toprule
 & \multicolumn{5}{c}{Dependent variable: No. of Abstract Views (in logarithm) by discipline} \\
 
 & {6 weeks} & {7 weeks} & {8 weeks} & {9 weeks} & {10 weeks} \\

{Variables} & {(1)} & {(2)} & {(3)} & {(4)}  & {(5)}  \\

\hline \\[-1.8ex]

{$Female$}                  & -0.083*    & -0.083*   & -0.083*    & -0.083*    & -0.083* \\
                            & {(0.047)}  & {(0.047)} & {(0.048)} & {(0.048)} & {(0.049)} \\ 
  
{$Female \times Lockdown$}  & 0.103      & 0.112     & 0.090     & 0.087     & 0.050 \\
                            & {(0.085)}  & {(0.079)} & {(0.074)} & {(0.070)} & {(0.068)} \\
  
{Time fixed effects}        & {Yes}      & {Yes}     & {Yes}     & {Yes}     & {Yes} \\
{Discipline fixed effects}  & {Yes}      & {Yes}     & {Yes}     & {Yes}     & {Yes} \\
Observations                & {720}      & {756}    & {792}      & {828}      & {864} \\
$R^{2}$                     & 0.828      & 0.829     & 0.830     & 0.831     & 0.834 \\
\bottomrule 
\end{tabular}
\noindent This table reports the estimated coefficients and robust standard errors (in parentheses) in Equation~\eqref{eq:did_view_download} using the discipline fixed effect, with $log(abstract\ views)$ as the dependent variable. The coefficients for 6, 7, 8, 9, and 10 weeks since the lockdown are presented in Columns (1)--(5), respectively. Significance at $^{*}p<0.1$; $^{**}p<0.05$; $^{***}p < 0.01$.
\end{threeparttable}
\end{table} 

\renewcommand{\arraystretch}{1.1}
\begin{table}[h] \centering \scriptsize \vspace{-0.1in}
 \caption{Impact of Lockdown on Downloads}
 \label{table:downloads_displine}
 \begin{threeparttable}
 \begin{tabular}{L{4cm} @{}l*{5}{S[table-format=-1.3,table-space-text-post=***,table-column-width=2cm]}@{}}
\toprule
 & \multicolumn{5}{c}{Dependent variable: No. of Downloads (in logarithm) by discipline} \\
 
 & {6 weeks} & {7 weeks} & {8 weeks} & {9 weeks} & {10 weeks} \\

{Variables} & {(1)} & {(2)} & {(3)} & {(4)}  & {(5)}  \\

\hline \\[-1.8ex]

{$Female$}                  & -0.049      & -0.049     & -0.049     & -0.049     & -0.044 \\
                            & {(0.042)}  & {(0.043)} & {(0.043)} & {(0.043)} & {(0.067)} \\ 
  
{$Female \times Lockdown$}  & -0.077      & -0.101     & -0.111     & -0.128*    & -0.131* \\
                            & {(0.085)}  & {(0.077)} & {(0.072)} & {(0.068} & {(0.065)} \\
  
{Time fixed effects}        & {Yes}      & {Yes}     & {Yes}     & {Yes}     & {Yes} \\
Observations                & {720}       & {756}      & {792}      & {828}      & {864} \\
$R^{2}$                     & 0.853      & 0.855     & 0.855     & 0.858     & 0.861 \\
\bottomrule 
\end{tabular}
\noindent This table reports the estimated coefficients and robust standard errors (in parentheses) in Equation~\eqref{eq:did_view_download} using the discipline level fixed effect, with $log(downloads)$ as the dependent variable. The coefficients for 6, 7, 8, 9, and 10 weeks since the lockdown are presented in Columns (1)--(5), respectively. Significance at $^{*}p<0.1$; $^{**}p<0.05$; $^{***}p < 0.01$.
\end{threeparttable}
\end{table} 

Tables~\ref{table:views_displine} and~\ref{table:downloads_displine} report the the effect of the lockdown on the number of abstract views and the number of downloads, respectively. In general, the average treatment effects are insignificant, suggesting that after the lockdown, female and male researchers' research quality did not change significantly, and that our findings are unlikely to be driven by shifts in research quality. In addition, column (4) and (5) of Table~\ref{table:downloads_displine} suggest that female researchers' research quality in terms of the number of downloads per preprint decreased compared to male researchers' after the lockdown. We repeat our analysis using Equation~\ref{eq:did} and find consistent results. The estimation results are reported in
Appendix Tables~\ref{table:did_views} and \ref{table:did_downloads}.

\subsection{Coauthorship}

We next study how coauthorship across genders affects our results. Recall that we measure the \textit{effective} productivity for preprints with $n$ authors by allocating $1/n$ preprint to each coauthor. That is, our measure implicitly assumes equal productivity across female and male authors. To alleviate the concern that this assumption may not hold, we conduct a sub-sample analysis focusing on preprints that have either \textit{all-male} or \textit{all-female} authors, excluding preprints that have both male and female authors. We repeat our DID analysis for this sub-sample using Equations~\eqref{eq:did_discipline} and~\eqref{eq:did} to compare the productivity gap between all-male and all-female preprints.


Table~\ref{table:did_allmf_bycat} reports the estimation results with the discipline fixed effect using Equation~\eqref{eq:did_discipline}. The results show that the number of all-female preprints has significantly dropped since the lockdown, compared to all-male preprints. Note that the coefficients of $Female \times Lockdown$ in Table~\ref{table:did_allmf_bycat} are more significant and larger than those in Table~\ref{table:did_discipline}, suggesting that gender inequality is more pronounced when a research team has only female authors. Intuitively, an all-female research team's capacity is more severely constrained, resulting in a lower productivity. Table~\ref{table:did_allmf_all} reports the results at the aggregate-level and the results are consistent. 


\renewcommand{\arraystretch}{1.1}
\begin{table}[h] \centering \scriptsize \vspace{-0.1in}
   \caption{Impact of Lockdown on Gender Inequality among All-male and All-female Preprints }
 \label{table:did_allmf_bycat}
 \begin{threeparttable}
 \begin{tabular}{ L{4cm} @{}c*{5}{S[table-format=-1.3,table-space-text-post=***,table-column-width=1.8cm]}@{}}
\toprule
 & \multicolumn{5}{c}{Dependent variable: No. of Preprints (in logarithm) by discipline} \\
 
 & {6 weeks} & {7 weeks} & {8 weeks} & {9 weeks} & {10 weeks} \\
 
{Variables} & {(1)} & {(2)} & {(3)} & {(4)}  & {(5)}  \\

\hline \\[-1.8ex]

{$Female$}  & $-$1.002*** & -1.002***  & -1.002*** & -1.002*** & -1.002*** \\
            & {(0.047)} & {(0.047)} & {(0.047)} & {(0.047)} & {(0.047)} \\ 
  
{$Female \times Lockdown$} &$-$0.233*** &  -0.263***  & -0.267***  & -0.255***  & -0.232*** \\
                           & {(0.088)} & {(0.084)} & {(0.079)} & {(0.076)} & {(0.074)} \\
 
Discipline fixed effects  & {Yes} & {Yes} & {Yes} & {Yes} & {Yes} \\
Time fixed effects & {Yes} & {Yes} & {Yes} & {Yes} & {Yes} \\
Observations & {720} & {756} & {792} & {828} & {864} \\
$R^{2}$ & 0.810 & 0.810 & 0.812 & 0.813 & 0.814 \\
\bottomrule 
\end{tabular}
\noindent This table reports the estimated coefficients and robust standard errors (in parentheses) in Equation~\eqref{eq:did_discipline} with the discipline level fixed effect. We restrict our sample to preprints that have either all-male authors, or all-female authors. The coefficients for 6, 7, 8, 9, and 10 weeks since the lockdown are presented in Columns (1)--(5), respectively.  Significance at $^{*}p<0.1$; $^{**}p<0.05$; $^{***}p < 0.01$.
\end{threeparttable}
\end{table} 

\subsection{Data Censoring}

One might be concerned that because research takes time, many researchers only post one preprint during our sample period. As a result, our data sample before and after the lockdown may contain a different set of authors. 

To address this concern, we perform two analyses. First, we collect additional data on US authors, extending the time window of our main analysis to 40 weeks. By adding 16 weeks to the post-treatment period, we capture more researchers in our sample, especially those who were not able to post preprints within 10 weeks of the lockdown. We repeat the DID analysis and find consistent results. 

Second, we construct a balanced panel by including only authors who posted preprints before the lockdown to compare the productivity of the same group of authors before and after the lockdown. This approach ensures an apple-to-apple comparison and helps us rule out potential biases introduced by different author samples in the pre-treatment and post-treatment periods. Table~\ref{table:balanced_aggregated} reports the estimated results using the balanced panel. The findings are consistent with our main results: within the same group of authors, female researchers' productivity dropped significantly after the lockdown compared to male researchers' productivity. 

\renewcommand{\arraystretch}{1.1}
\begin{table}[h] \centering \scriptsize \vspace{-0.1in}
   \caption{Impact of Lockdown on Gender Inequality Using the Balanced Panel}
 \label{table:balanced_aggregated}
 \begin{threeparttable}
 \begin{tabular}{ L{4cm} @{}c*{5}{S[table-format=-1.3,table-space-text-post=***,table-column-width=1.8cm]}@{}}
\toprule
 & \multicolumn{5}{c}{Dependent variable: No. of Preprints (in logarithm) in aggregation} \\
 
 & {6 weeks} & {7 weeks} & {8 weeks} & {9 weeks} & {10 weeks} \\
 
{Variables} & {(1)} & {(2)} & {(3)} & {(4)}  & {(5)}  \\

\hline \\[-1.8ex]

{$Female$}  & $-$0.920*** & -0.920***  & -0.920*** & -0.920*** & -0.920*** \\
            & {(0.065)} & {(0.065)} & {(0.065)} & {(0.065)} & {(0.065)} \\ 
  
{$Female \times Lockdown$} &$-$0.518*** &  -0.638***  & -0.589***  & -0.556***  & -0.641*** \\
                           & {(0.138)} & {(0.170)} & {(0.159)} & {(0.148)} & {(0.159)} \\
 
Discipline fixed effects  & {Yes} & {Yes} & {Yes} & {Yes} & {Yes} \\
Time fixed effects & {Yes} & {Yes} & {Yes} & {Yes} & {Yes} \\
Observations & {40} & {42} & {44} & {46} & {48} \\
$R^{2}$ & 0.991 & 0.988 & 0.988 & 0.988 & 0.985 \\
\bottomrule 
\end{tabular}
\noindent This table reports the estimated coefficients and robust standard errors (in parentheses) in Equation~\eqref{eq:did} at the aggregate level. The coefficients for 6, 7, 8, 9, and 10 weeks since the lockdown are presented in Columns (1)--(5), respectively. Significance at $^{*}p<0.1$; $^{**}p<0.05$; $^{***}p < 0.01$.
\end{threeparttable}
\end{table}

\section{Conclusions}
Our paper adds to the long-standing literature on gender equality, an important topic in social science. For example, the literature has shown evidence of fairness in parental leaves \citep{lundquist2012parental} and inequality in tenure evaluation \citep{sarsons2017recognition,antecol2018equal}, recognition \citep{ghiasi2015compliance}, and compensation \citep{pierce2020peer}. Researchers have also investigated business innovations to help empower women \citep{l2020alleviating}. The COVID-19 crisis brings to the forefront a long existing issue: the inequities faced by women who often do more of the childcare and housework. 
We contribute to the literature by providing direct tests of the impact of the pandemic shock on gender inequality in academia.



We show that in the US, since the lockdown began, women have produced 13.2--13.9-percent fewer social science research papers than men. We also find that the effect is especially significant for junior faculty and for researchers at top-ranked universities. Finally, we find that the increase in productivity inequality is significant in seven countries. The results are robust when we repeat our analysis over papers with same-gender authors, a balanced sample with the same group of authors before and after the lockdown, and an extended data sample. 

Our findings indicate that, if the lockdown is kept in place for too long, female academics in junior positions and at top-ranked universities are likely to be significantly disadvantaged---a fairness issue that may expose women to a higher unemployment or career risk in the future. We hope our findings increase awareness of this issue. Actions could be taken to balance domestic responsibilities among spouses. Recently, many universities have taken actions such as granting tenure clock extensions to both female and male faculty. Recall that our paper finds an overall 35-percent increase in productivity and a 13-percent increase in gender gap among social science researchers. Therefore, our findings do not provoke a concern for overall productivity but rather for gender inequality. As a result, universities could consider providing additional support, such as childcare support, to female researchers whose productivity has been disproportionately affected. Universities and letter writers should keep this inequality in mind when evaluating professors for promotion. We also hope our work will inspire researchers to explore other forms of inequality arising from the COVID-19 pandemic. 

Our findings also suggest that telecommuting may have unintended consequences for gender inequality. As the COVID-19 outbreak accelerates the trend toward telecommuting, institutions and firms should take gender equality into consideration when designing and implementing telecommuting policies. We hope that our work could serve as a stepping stone to stimulate more research on the synergy between operations and social issues. 

Our study has a few limitations. First, since it focuses on social science disciplines, and thus the findings may not be generalizable to other disciplines. Second, we have limited information about the researchers in our dataset. Future research could collect additional data---such as parental status, whether they are allocating more or less time to research than they did before the pandemic, whether they multitask at home, and who performs household duties---to pinpoint the exact mechanism underlying the observed empirical patterns. 



\newpage
\bibliographystyle{ormsv080}
\addcontentsline{toc}{section}{\refname}
\bibliography{main}

\newpage

\renewcommand\thefigure{A.\arabic{figure}} 
\renewcommand\thetable{A.\arabic{table}} 
\setcounter{figure}{0}
\setcounter{table}{0}
\pagenumbering{roman}

\begin{APPENDIX}{}

\begin{figure}[h]
\caption{Time Trends of US Preprints from December 2018 to May 2019}
\label{fig:trend2019}
\vspace{0.05in}
\centering
\includegraphics[scale=0.36]{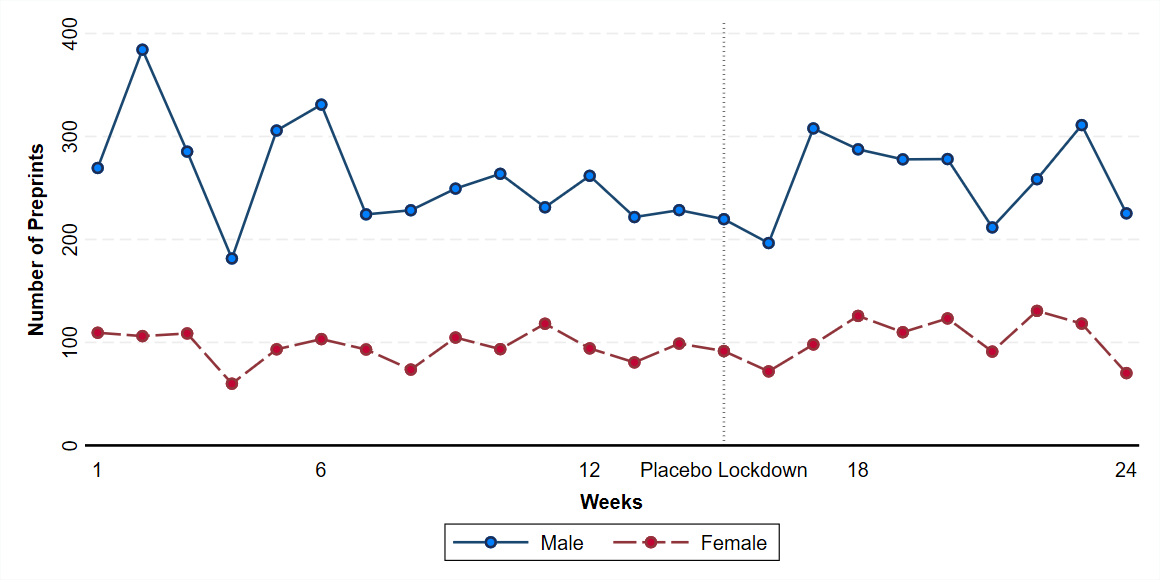}
\vspace{0.05in}

\noindent \scriptsize This graph plots the time trend of the number of preprints for female academics and male academics. The vertical line represents the placebo lockdown week (the week of March 11) in 2019.

\end{figure}

\renewcommand{\arraystretch}{1.1}
\begin{table}[h] \centering \scriptsize \vspace{-0.1in}
   \caption{Robustness to Different University Rankings}
 \label{table:rankingrobust}
 \begin{threeparttable}
\begin{tabular}{ @{}l*{5}{S[table-format=-1.3,table-space-text-post=0***,table-column-width=1.8cm]}@{}}
\toprule
 & \multicolumn{5}{c}{Dependent variable: No. of Preprints  (in logarithm) by discipline} \\
{Universities} & {6 weeks} & {7 weeks} & {8 weeks} & {9 weeks} & {10 weeks} \\
 
 
by Times ranking & {(1)} & {(2)} & {(3)} & {(4)}  & {(5)}  \\

\hline \\[-1.8ex]

Top 10  & -0.209*** & -0.230***  & -0.198*** & -0.185*** &  -0.181*** \\

Top 20  & -0.177** & -0.222***  & -0.205*** & -0.204***&  -0.214*** \\

Top 30  & -0.227*** & -0.253***  & -0.228*** & -0.228*** &  -0.228*** \\

Top 40  & -0.157** & -0.211***  & -0.196*** & -0.196***&  -0.202*** \\

Top 50  & -0.114 & -0.147**  & -0.130* & -0.138** &  -0.146** \\

Top 60  & -0.126* & -0.143*  & -0.131* & -0.137** &  -0.147** \\

Top 70  & -0.142* & -0.157**  & -0.141** & -0.143** &  -0.143** \\

Top 80  & -0.139* & -0.154**  & -0.140** & -0.131* &  -0.130** \\

Top 90  & -0.134* & -0.146**  & -0.137** & -0.133* &  -0.135** \\

Top 100  & -0.124 & -0.129*  & -0.125* & -0.118* &  -0.118* \\

Observations & {720} & {756} & {792} & {828} & {864}\\

\midrule

& \multicolumn{5}{c}{Dependent variable: No. of preprints  (in logarithm) by discipline} \\
 
Universities  & {6 weeks} & {7 weeks} & {8 weeks} & {9 weeks} & {10 weeks}   \\
by ARWU ranking & {(1)} & {(2)} & {(3)} & {(4)}  & {(5)}  \\

\hline \\[-1.8ex]

Top 10  & -0.232*** & -0.255***  & -0.233*** & -0.214*** &  -0.222*** \\

Top 20  & -0.259** & -0.297***  & -0.271*** & -0.260***&  -0.256*** \\

Top 30  & -0.261*** & -0.305***  & -0.268*** & -0.264*** &  -0.259*** \\

Top 40  & -0.136* & -0.188**  & -0.171** & -0.176*** &  -0.171*** \\

Top 50  & -0.104 & -0.156**  & -0.132* & -0.133** &  -0.139** \\

Top 60  & -0.171** & -0.154***  & -0.154*** & -0.143***&  -0.114* \\

Top 70  & -0.080 & -0.125*  & -0.109 & -0.113* &  -0.120* \\

Top 80  & -0.123 & -0.128*  & -0.117* & -0.118* &  -0.120* \\

Top 90  & -0.099 & -0.105  & -0.095 & -0.093 &  -0.096 \\

Top 100  & -0.090 & -0.094  & -0.086 & -0.084&  -0.089 \\

Observations & {720} & {756} & {792} & {828} & {864} \\

\bottomrule 
\end{tabular}

\noindent This table reports the estimated coefficients in Equation~\eqref{eq:did} across universities with different rankings. The coefficients for 6, 7, 8, 9 and 10 weeks since the lockdown are presented in columns (1)--(5), respectively. Time fixed effects at the weekly level are included in all regressions. Note that we omit reporting standard errors and estimates of other variables for brevity. Significance at $^{*}p<0.1$; $^{**}p<0.05$; $^{***}p < 0.01$.
\end{threeparttable}
\vspace{-0.1in}
\end{table}

\renewcommand{\arraystretch}{1.1}
\begin{table}[h] \centering \scriptsize \vspace{-0.1in}
 \caption{Parallel Trends Test}
 \label{table:trend}
 \begin{threeparttable}
\begin{tabular}{@{}l *{2}{S[table-format=+1.5,table-space-text-post=***,parse-numbers = false]}@{} }
\toprule
 
 & {No. of Preprints  (in logarithm) in aggregation} & {No. of Preprints  (in logarithm) by discipline}   \\
Variables & $\text{ }\text{ }{(1)}$ & $\text{ }\text{ }{(2)}$ \\ \hline \\[-1.8ex]

 {$Female \times Time_{-14}$} & -0.231 & -0.189  \\
& (0.430) & (0.352) \\

 {$Female \times Time_{-13}$} & -0.013 & 0.157  \\
& (0.430) & (0.335) \\

 {$Female \times Time_{-12}$} & -0.377 & -0.202  \\
& (0.430) & (0.309) \\

 {$Female \times Time_{-11}$} & 0.060 & 0.219 \\
& (0.430) & (0.302) \\

 {$Female \times Time_{-10}$} & -0.030 & -0.054  \\
& (0.430) & (0.210) \\

 {$Female \times Time_{-9}$} & -0.028 & -0.213  \\
& (0.430) & (0.243) \\

{$Female \times Time_{-8}$} & -0.144 & -0.146  \\
& (0.430) & (0.258) \\

{$Female \times Time_{-7}$} & -0.101 & -0.031  \\
& (0.430) & (0.234) \\

{$Female \times Time_{-6}$} & -0.363 & -0.413$**$  \\
& (0.430) & (0.250) \\

{$Female \times Time_{-5}$} & 0.355 & 0.314$*$ \\
& (0.430) & (0.214) \\
{$Female \times Time_{-4}$} & 0.130 & 0.063  \\
& (0.430) & (0.224) \\
{$Female \times Time_{-3}$} & 0.098 & -0.051  \\
& (0.430) & (0.218) \\
{$Female \times Time_{-2}$} & 0.069 & 0.056  \\
& (0.430) & (0.239) \\
{$Female \times Time_{-1}$} & 0.092 & 0.190   \\
& (0.430) & (0.219) \\

{Observations} & \text{ }\text{ }{24} & \text{ }\text{ }{540} \\
{$R^{2}$} & 0.894 & 0.808  \\

\bottomrule 
\end{tabular}
\noindent This table reports the estimated coefficients of the parallel trends test using Equation~\eqref{eq:did_parallel}. The results at the aggregate-level and discipline-level are presented in Columns (1) and (2), respectively. Note that we omit reporting estimates of other variables for brevity. Time fixed effects at the weekly level are included in all regressions. Significance at $^{*}p<0.1$; $^{**}p<0.05$; $^{***}p < 0.01$.
\end{threeparttable}
\vspace{-0.1in}
\end{table}

\renewcommand{\arraystretch}{1.1}
\begin{table}[h] \centering \scriptsize \vspace{-0.1in}
   \caption{Impact of Lockdown on Gender Inequality by Academic Ranks in Aggregation}
 \label{table:rank_aggregated}
 \begin{threeparttable}
\begin{tabular}{  @{}l*{5}{S[table-format=-1.3,table-space-text-post=***,table-column-width=1.8cm]}@{}}
\toprule

 & \multicolumn{5}{c}{Dependent variable: No. of Preprints (in logarithm) in aggregation} \\

{Researchers} & {6 weeks} & {7 weeks} & {8 weeks} & {9 weeks} & {10 weeks} \\
 
{by Academic Ranks} & {(1)} & {(2)} & {(3)} & {(4)}  & {(5)}  \\

\hline \\[-1.8ex]

Student & -0.030 &  -0.042  & -0.039  & -0.061  & -0.091 \\
Assistant Prof. & -0.485*** & -0.384** & -0.408** & -0.383** & -0.405** \\
Associate Prof. & 0.046 & -0.013 & 0.016 & -0.029 & -0.019 \\
Full Prof. & -0.175 & -0.154 & -0.055 & -0.035 & -0.052 \\

Observations & {40} & {42} & {44} & {46} & {48} \\
\bottomrule 
\end{tabular}
\noindent  This table reports the estimated coefficients based on Equation~\eqref{eq:did} for researchers within each rank group. The coefficients for 6, 7, 8, 9, and 10 weeks since the lockdown are presented in columns (1)--(5), respectively. Time fixed effects at the weekly level are included in each regression. Standard errors and estimates of other variables are omitted for brevity. Significance at $^{*}p<0.1$; $^{**}p<0.05$; $^{***}p < 0.01$.
\end{threeparttable}
\end{table} 

\renewcommand{\arraystretch}{1.1}
\begin{table}[h] \centering \scriptsize \vspace{-0.1in}
   \caption{Impact of Lockdown on Gender Inequality among All Male or All Female Preprints in Aggregation}
 \label{table:did_allmf_all}
 \begin{threeparttable}
 \begin{tabular}{ L{4cm} @{}c*{5}{S[table-format=-1.3,table-space-text-post=***,table-column-width=2cm]}@{}}
\toprule
 & \multicolumn{5}{c}{Dependent variable: No. of Preprints (in logarithm) in aggregation} \\
 
 & {6 weeks} & {7 weeks} & {8 weeks} & {9 weeks} & {10 weeks} \\
 
{Variables} & {(1)} & {(2)} & {(3)} & {(4)}  & {(5)}  \\

\hline \\[-1.8ex]

{$Female$}  & $-$1.253*** & -1.253***  & -1.253*** & -1.253*** & -1.253*** \\
            & {(0.066)} & {(0.066)} & {(0.066)} & {(0.066)} & {(0.065)} \\ 
  
{$Female \times Lockdown$} &$-$0.285** &  -0.297***  & -0.254**  & -0.233**  & -0.220** \\
                           & {(0.099)} & {(0.092)} & {(0.096)} & {(0.093)} & {(0.089)} \\
 
Discipline Fixed Effects  & {Yes} & {Yes} & {Yes} & {Yes} & {Yes} \\
Time Fixed Effects & {Yes} & {Yes} & {Yes} & {Yes} & {Yes} \\
Observations & {40} & {42} & {44} & {46} & {48} \\
$R^{2}$ & 0.978 & 0.979 & 0.978 & 0.978 & 0.979 \\
\bottomrule 
\end{tabular}
\noindent This table reports the estimated coefficients and robust standard errors (in parentheses) in Equation~\eqref{eq:did} at the aggregate level. We restrict our sample to those preprints that have either all-male authors, or all-female authors. The coefficients for 6, 7, 8, 9, and 10 weeks since the lockdown are presented in columns (1)--(5), respectively. Significance at $^{*}p<0.1$; $^{**}p<0.05$; $^{***}p < 0.01$.
\end{threeparttable}
\end{table}

\begin{table}[h]\centering \scriptsize
\caption{Summary Statistics for December 2018 - May 2019} 
\label{table:statistics2}
 \begin{threeparttable}
\begin{tabular}{C{1.4cm} l*{9}{S[table-format=4.2, table-number-alignment=center]}}

\toprule
 &  &  \multicolumn{5}{c}{All observations} & \multicolumn{2}{c}{Before March 2019} & \multicolumn{2}{c}{After March 2019} \\
 \cmidrule(lr){3-7} \cmidrule(lr){8-9} \cmidrule(lr){10-11} 

Level  &  \centering{Weekly No. of Preprints} &  {Mean}  & {Std. dev}  & {Max} & {Min} & {Total} &  {Mean}  & {Std. dev} &  {Mean}  & {Std. dev} \\

\midrule

\multirow{3}{1.4cm}{\centering All Disciplines (US only)} 
& All  & 401.0 & 69.6 & 535 & 267 & {9,333} & 406.4 & 75.8 & 393.3 & 58.9 \\
& Female authors & 103.0 & 17.2 & 131 & 62 &  {2,413}  & 102.1 & 15.1 & 104.4 & 19.7 \\
& Male authors  & 298.0 & 57.9 & 424 & 205 & {6,920} & 304.3 & 65.7 & 288.9 & 42.7\\
\hline \\[-1.8ex]

\multirow{18}{1.4cm}{\centering By Discipline (US only)} 
& Accounting    & 21.0 & 6.3 & 34 & 10 & {505} & 21.9 & 6.6 & 19.9 & 6.2 \\
& Anthropology  & 76.3 & 19.9 & 115 & 41 & {1,832} & 69.4 & 20.9 & 86.1 & 14.0\\
& Cognitive    & 17.0 & 7.7 & 38 & 7 & {407} & 20.5 & 7.9 & 12.0 & 3.7 \\
& Corporate & 17.5 & 5.9 & 30 & 8 & {420} & 17.2 & 5.6 & 17.9 & 6.4 \\
& Criminal & 16.3 & 5.6 & 32 & 6 & {390} & 14.9 & 6.4 & 18.2 & 3.8 \\
& Economics & 212.0 & 50.9 & 348 & 133 & {5,089} & 225.7 & 55.7 & 192.9 & 37.9 \\
& Education & 15.3 & 5.2 & 29 & 6 & {366} & 15.3 & 5.2 & 15.2 & 5.6 \\
& Entrepreneurship & 16.1 & 5.6 & 28 & 8 & {387} & 18.7 & 5.3 & 12.5 & 3.6 \\
& Finance & 89.7 & 21.3 & 148 & 66 & {2,153} & 95.0 & 25.2 & 82.3 & 11.8 \\
& Geography & 13.6 & 6.3 & 29 & 5 & {327} & 11.9 & 4.9 & 16.0 & 7.5 \\
& Health Economics & 4.3 & 4.2 & 22 & 0 & {104} & 3.3 & 1.7 & 5.8 & 6.1 \\
& Information Systems & 20.2 & 5.8 & 36 & 10 & {485} & 22.0 & 6.4 & 17.7 & 3.9 \\
& Law & 143.1 & 32.6 & 211 & 76 & {3,434} & 135.4 & 36.3 & 153.8 & 24.4 \\
& Management & 32.4 & 11.8 & 57 & 8 & {778} & 34.7 & 11.1 & 29.2 & 12.5 \\
& Organization & 24.8 & 7.8 & 43 & 15 & {594} & 27.2 & 8.4 & 21.3 & 5.7 \\
& Political Science & 166.3 & 28.3 & 225 & 124 & {3,991} & 172.5 & 30.9 & 157.6 & 22.8 \\
& Sustainability & 38.8 & 23.9 & 105 & 14 & {930} & 34.1 & 16.7 & 45.2 & 31.3 \\
& Women/Gender & 19.4 & 8.4 & 40 & 4 & {466} & 20.9 & 9.9 & 17.4 & 5.8 \\


\bottomrule
\end{tabular}

\noindent The table summarizes the weekly number of papers from December 2018 to May 2019. In total,  there are 9,333 preprints produced by 14,767 US authors, 2,413 of which are produced by 3,876 female researchers and 6,920 are produced by 10,891 male researchers. We gather the country-specific lockdown time to split our sample to before and after the lockdown for each country.

\end{threeparttable}
\end{table}

\renewcommand{\arraystretch}{1.1}
\begin{table}[h] \centering \scriptsize \vspace{-0.1in}
  \caption{Falsification Test}
 \label{table:falsification}
 \begin{threeparttable}
 \begin{tabular}{ @{}l*{5}{S[table-format=-1.3,table-space-text-post=0***,table-column-width=1.8cm]}@{}}
\toprule
 & \multicolumn{5}{c}{Dependent variable: No. of Preprints  (in logarithm) in aggregation} \\
 & {6 weeks} & {7 weeks} & {8 weeks} & {9 weeks} & {10 weeks} \\
 & {(1)} & {(2)} & {(3)} & {(4)}  & {(5)}  \\
\hline \\[-1.8ex]
{$Female \times Lockdown$}  & 0.042 & 0.061 & 0.088 & 0.080 &  0.057 \\
{Observations} & {40} & {42} & {44} & {46} & {48} \\
{$R^2$} & 0.980 & 0.980 & 0.979 & 0.980 & 0.980\\

\midrule
 & \multicolumn{5}{c}{Dependent variable: No. of Preprints  (in logarithm) by discipline} \\
{$Female \times Lockdown$}  & 0.092 & 0.094  & 0.103* & 0.085 &  0.070 \\
{Observations} & {720} & {756} & {792} & {828} & {864} \\
{$R^2$} & 0.877 & 0.877 & 0.871 & 0.873 & 0.873\\

\bottomrule 
\end{tabular}
\noindent This table reports the estimated coefficients of the interacted term, Female $\times$ Lockdown, in Equation~\eqref{eq:did}. The coefficients for 6, 7, 8, 9 and 10 weeks since the lockdown are presented in columns (1)--(5), respectively. Note that we omit reporting estimates of other variables for brevity. Time fixed effects at the weekly level are included in all regressions. Significance at $^{*}p<0.1$; $^{**}p<0.05$; $^{***}p < 0.01$.
\end{threeparttable}
\end{table} 

\begin{table}[h]\centering \scriptsize
    \caption{Summary Statistics for Downloads and Abstract Views} 
    \label{table:statistics_downloads_views}
    \begin{threeparttable}
    \begin{tabular}{C{2cm} l*{8}{S[table-format=4.2, table-number-alignment=center]}}

\toprule
 &  &  \multicolumn{4}{c}{All observations} & \multicolumn{2}{c}{Before Lockdown} & \multicolumn{2}{c}{After Lockdown} \\
 \cmidrule(lr){3-6} \cmidrule(lr){7-8} \cmidrule(lr){9-10} 

Level  &  \centering{Groups} &  {Mean}  & {Std. dev}  & {Min} & {Max} &  {Mean}  & {Std. dev} &  {Mean}  & {Std. dev} \\

\midrule

\multirow{3}{2cm}{\centering No. of downloads per preprint} 
& All  & 40.9 & 18.6 & 13.7 & 84.6 & 54.2 & 14.6 & 26.9 & 10.2 \\
& Female authors & 39.2 & 21.1 & 10.2 & 85.6 & 53.0 & 18.1 & 24.6 & 12.5 \\
& Male authors  & 41.7 & 18.6 & 14.9 & 84.2 & 54.8 & 14.6 & 27.8 & 10.7\\
\hline \\[-1.8ex]

\multirow{3}{2cm}{\centering  No. of abstract views per preprint} 
& All  & 144.5 & 47.6 & 57.67 & 226.1  & 184.0 & 19.4 & 102.7 & 29.2 \\
& Female authors & 139.1 & 49.0 & 44.7 & 243.1 & 176.4 & 26.6 & 99.6 & 34.1 \\
& Male authors & 146.8 & 48.2 & 62.1 & 232.3 & 187.1 & 19.8 & 104.1 & 28.6 \\

\bottomrule
\end{tabular}

\noindent The table summarizes the weekly average number of downloads and abstract views per preprint from December 2019 to May 2020. The sample includes 9,934 preprints from authors in the United States.

\end{threeparttable}
\end{table}

\renewcommand{\arraystretch}{1.1}
\begin{table}[h] \centering \scriptsize \vspace{-0.1in}
 \caption{Impact of Lockdown on Abstract Views}
 \label{table:did_views}
 \begin{threeparttable}
 \begin{tabular}{L{4cm} @{}l*{5}{S[table-format=-1.3,table-space-text-post=***,table-column-width=2cm]}@{}}
\toprule
 & \multicolumn{5}{c}{Dependent variable: No. of Abstract Views (in logarithm) in aggregation} \\
 
 & {6 weeks} & {7 weeks} & {8 weeks} & {9 weeks} & {10 weeks} \\

{Variables} & {(1)} & {(2)} & {(3)} & {(4)}  & {(5)}  \\

\hline \\[-1.8ex]

{$Female$}                  & -0.054      & -0.054     & -0.054     & -0.054     & -0.054 \\
                            & {(0.048)}  & {(0.048)} & {(0.048)} & {(0.048)} & {(0.048)} \\ 
  
{$Female \times Lockdown$}  & 0.086      & 0.088     & 0.074     & 0.067     & 0.044 \\
                            & {(0.074)}  & {(0.068)} & {(0.065)} & {(0.062)} & {(0.058)} \\
  
{Time Fixed Effects}        & {Yes}      & {Yes}     & {Yes}     & {Yes}     & {Yes} \\
Observations                & {40}       & {42}      & {44}      & {46}      & {48} \\
$R^{2}$                     & 0.894      & 0.913     & 0.935     & 0.948     & 0.955 \\
\bottomrule 
\end{tabular}
\noindent This table reports the estimated coefficients and robust standard errors (in parentheses) in Equation~\eqref{eq:did}, with \textit{abstract views} as the dependent variable. The coefficients for 6, 7, 8, 9 and 10 weeks since the lockdown are presented in columns (1)--(5), respectively. Significance at $^{*}p<0.1$; $^{**}p<0.05$; $^{***}p < 0.01$.
\end{threeparttable}
\end{table} 

\renewcommand{\arraystretch}{1.1}
\begin{table}[h] \centering \scriptsize \vspace{-0.1in}
 \caption{Impact of Lockdown on Downloads}
 \label{table:did_downloads}
 \begin{threeparttable}
 \begin{tabular}{L{4cm} @{}l*{5}{S[table-format=-1.3,table-space-text-post=***,table-column-width=2cm]}@{}}
\toprule
 & \multicolumn{5}{c}{Dependent variable: No. of Downloads (in logarithm) in aggregation} \\
 
 & {6 weeks} & {7 weeks} & {8 weeks} & {9 weeks} & {10 weeks} \\

{Variables} & {(1)} & {(2)} & {(3)} & {(4)}  & {(5)}  \\

\hline \\[-1.8ex]

{$Female$}                  & -0.044      & -0.044     & -0.044     & -0.044     & -0.044 \\
                            & {(0.067)}  & {(0.067)} & {(0.067)} & {(0.067)} & {(0.067)} \\ 
  
{$Female \times Lockdown$}  & -0.027      & -0.057     & -0.068     & -0.085    & -0.087 \\
                            & {(0.175)}  & {(0.157)} & {(0.141)} & {(0.130)} & {(0.120)} \\
  
{Time Fixed Effects}        & {Yes}      & {Yes}     & {Yes}     & {Yes}     & {Yes} \\
Observations                & {40}       & {42}      & {44}      & {46}      & {48} \\
$R^{2}$                     & 0.836      & 0.866     & 0.891     & 0.910     & 0.927 \\
\bottomrule 
\end{tabular}
\noindent This table reports the estimated coefficients and robust standard errors (in parentheses) in Equation~\eqref{eq:did}, with \textit{downloads} as the dependent variable. The coefficients for 6, 7, 8, 9 and 10 weeks since the lockdown are presented in columns (1)--(5), respectively. Significance at $^{*}p<0.1$; $^{**}p<0.05$; $^{***}p < 0.01$.
\end{threeparttable}
\end{table}

\end{APPENDIX}

\end{document}